\documentclass[%
 reprint,
 amsmath,amssymb,
 aps,
prb]{revtex4-1}

\usepackage{color}
\usepackage{subfigure}
\usepackage{graphicx}
\usepackage{dcolumn}
\usepackage{bm}

\begin{document}

\preprint{APS/123-QED}

\title{Nonreciprocal superposition state in antiferromagnetic optospintronics}

\author{Nobuyuki Okuma}
\email{okuma@hosi.phys.s.u-tokyo.ac.jp}
\affiliation{%
 Yukawa Institute for Theoretical Physics, Kyoto University, Kyoto 606-8502, Japan
}%

%
\date{\today}
\begin{abstract}

The absence of net magnetization, which forbids any stray magnetic fields, is one of the greatest advantages of antiferromagnets in device applications.
In conventional antiferromagnets, however, spin current cannot be extracted without the aid of a static magnetic field.
Here, we develop a theory of antiferromagnetic opto-spintronics to resolve this fundamental dilemma.
By coupling a linearly polarized photon and nonreciprocal magnon bands, we construct a superposition state of left- and right-handed magnon states with opposite group velocities.
We numerically demonstrate that by using this superposition state, an antiferromagnetic spin current can be efficiently generated without a net magnetic field including net magnetization.
We also find that the breakdown of the superposition state induces the stripe superfluid phase of a two-component Bose-Einstein condensate.
Our results lay the foundation for manipulating the superposition states of emergent particles in devices.
\end{abstract}

\maketitle


\section{Introduction}
Magnons, the quanta of spin wave fluctuations around ordered magnetic states, have attracted much interest in modern condensed matter physics \cite{magnonics, chumak}.
Owing to their long lifetime \cite{cornelissen} and finite spin angular momentum, magnons as well as electrons are important carriers in device applications.
In particular, antiferromagnetic magnons have, in addition to their ultrafast nature \cite{jungwirth,optospintronics}, a degenerate spin degree of freedom that can be understood as an analogue of photon polarization \cite{rcheng,antiferrospinpump1,igor}.
Recent studies have proposed methods for controlling magnon polarization by using an electric field \cite{rcheng} or circularly polarized light \cite{antiferrospinpump1,tsatoh}.

In terms of spin transport, however, conventional antiferromagnets present a fundamental dilemma.
The absence of net magnetization, which forbids any stray magnetic fields in the system, is a great advantage of antiferromagnets \cite{jungwirth}.
However, a spin current cannot be extracted \cite{ferri} without applying a static magnetic field to split the magnon spin degeneracy or allowing a net magnetization in antiferromagnets, which means that the feature of zero net magnetization is not helpful.

Band engineering is often a good solution for such a fundamental problem in condensed matter physics.
As in the case of electron systems, one can construct profound band structures by using exotic lattice structures, noncollinear magnetic orders, the Dzyaloshinskii-Moriya (DM) interaction, and so on.
Numerous concepts in multiband electron systems have been generalized to magnonic systems, e.g., the magnon Hall effect \cite{fujimoto,katsura,onose,owerre3}, spin-momentum locking \cite{okuma,kawano}, topological insulators \cite{shindou,lzhang,chisnell,owerre}, and topological semimetals \cite{fyli,jianandnie,fransson,pershoguba,okuma,nodalloop,nodalloop3,nodalloop2}.  
Among these concepts, one of the simplest nontrivial examples is the nonreciprocal magnon band \cite{ferrononreciprocal} in antiferromagnets \cite{okuma,hayami,rcheng,noncentrosym,takashima}.
In the presence of the DM interaction and easy-axis anisotropy, two branches, both of which are asymmetric with respect to momentum, appear in the magnon band structure.
This band structure has been extensively investigated theoretically \cite{okuma,hayami,rcheng} and experimentally \cite{noncentrosym,takashima}, including direct observation of the band structure in neutron scattering measurements \cite{noncentrosym}.

\begin{figure}[]
\begin{center}
　　　\includegraphics[width=7cm,angle=0,clip]{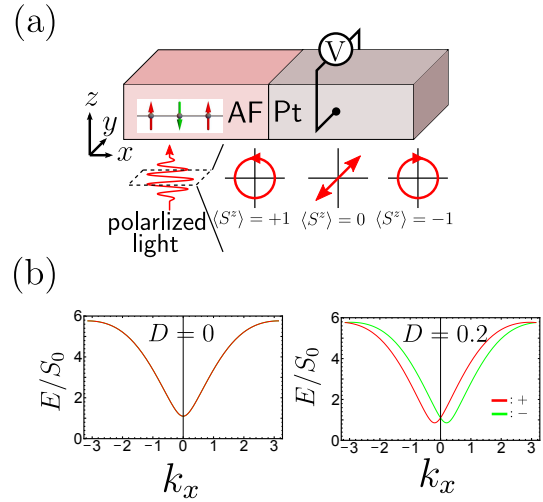}
　　　\caption{(a) Experimental setup of antiferromagnetic (AF) spin pumping under polarized light. The $z$-component spin current is detected as the $y$-direction inverse spin Hall voltage in the attached metal (Pt). Polarization-dependent and polarization-independent contributions arise from diffusive and drift spin currents in the presence of the DM interaction (see the main text for details). (b) Magnon dispersion in the $k_x$ direction for $D=0$ and $D=0.2$ with $J=1$, $K=-0.05$, and $k_y=k_z=0$. The band indices $\pm$ correspond to the spin angular momentum $\pm1$.}
　　　\label{fig1}
\end{center}
\end{figure}

In this paper, we present a theory of nonreciprocal magnons coupled with polarized photons in order to resolve the dilemma of spin transport in antiferromagnets.
In particular, by using linearly polarized light, we construct a superposition state of left- and right-handed magnon states with opposite group velocities, which carry a drift spin current without net particle or thermal currents.
We numerically demonstrate that this superposition state can be used for efficient spin pumping in an actual antiferromagnet without generating any net static magnetic fields, including net magnetization. 
We also find that the spin oscillation observed in numerical simulation is a precursory phenomenon of the stripe superfluid, which has been recently discussed for ultracold atomic systems.
These results are manifestations of observable phenomena induced by the spin-momentum-locked magnon band in the presence of the DM interaction.

This paper is organized as follows. 
In Sec. \ref{sec2}, we develop a quantum theory of reciprocal and nonreciprocal magnons coupled with polarized photons.
Using this quantum description, we qualitatively propose a theory of nonreciprocal spin pumping realized in a magnonic superposition state.
 In Sec. \ref{sec3}, we perform numerical simulations, based on the Landau-Lifshitz-Gilbert equation, to estimate quantitatively the nonreciprocal spin current. The spin density distribution of the superposition state in real space is also discussed.
We relate the spin oscillation observed in a numerical simulation with the two-component Bose-Einstein condensate in Sec. \ref{sec4}.

\section{Quantum theory of nonreciprocal spin pumping\label{sec2}}
In this section, we quantize the polarized light and spin fluctuation in antiferromagnets with and without the DM interaction in order to discuss the perfect interconversion of polarization. We first ignore the presence of Pt and consider the infinite-size system with the periodic boundary conditions.
We then propose the nonreciprocal spin pumping in the presence of Pt qualitatively.
In the following, we set $\hbar=1$.

\subsection{Quantum description of polarization conversion}
In this paper, we consider a longitudinal spin pumping measurement in which a collinear antiferromagnet and metal (Pt) are used as the spin current generator and detector, respectively [Fig. $\ref{fig1}$(a)]. Spin excitations are resonantly excited by polarized light.
The collinear antiferromagnetic order is set to be parallel to the photon propagation direction (the $z$ direction).
Injected $z$-component spin current is detected as the $y$-direction inverse spin Hall voltage.
In the following, we set $\hbar=1$.

We first review the spin dynamics in an antiferromagnet with the DM interaction\cite{okuma,hayami,rcheng,noncentrosym,takashima} caused by the inversion breaking in the $x$ direction \cite{orientation}:
\begin{align}
H_{spin}=\sum_{\langle \bm{i},\bm{j} \rangle}\left[J\bm{S}_{\bm{i}}\cdot\bm{S}_{\bm{j}}+\bm{D}\cdot\left(\bm{S}_{\bm{i}}\times\bm{S}_{\bm{j}} \right)\right]+K\sum_{\bm{i}}[S^z_{\bm{i}}]^2,\label{1dantiferro}
\end{align}
where $J>0$ is the nearest-neighbour antiferromagnetic exchange coupling and $\bm{D}=D\hat{z}\delta_{i_yj_y}\delta_{i_zj_z}$ denotes the $z$-component DM interaction in the $x$ direction. 
The easy-axis anisotropy $K<0$ is set to be large enough to stabilize the collinear antiferromagnetic ground state with two sublattices A and B (see Appendix).

To correctly treat spin angular momentum carried by one quantum, we quantize the spin wave fluctuation.
In the semiclassical picture, spin excitations around the ground state are approximated by Holstein-Primakoff bosons ($c,c^{\dagger}$):
\begin{align}
&S^{\pm}_{\bm{R},A}\simeq\sqrt{2S_0}c^{(\dagger)}_{\bm{R},A}, S^z_{\bm{R},A}=S_0-c^{\dagger}_{\bm{R},A}c_{\bm{R},A},\notag\\
&S^{\mp}_{\bm{R},B}\simeq\sqrt{2S_0}c^{(\dagger)}_{\bm{R},B}, S^z_{\bm{R},B}=c^{\dagger}_{\bm{R},B}c_{\bm{R},B}-S_0,\label{holstein}
\end{align}
where $S_0=|\bm{S}|$ and $\bm{R}$ denotes the magnetic lattice vector.
Using Eq. ($\ref{holstein}$), we can rewrite Eq. ($\ref{1dantiferro}$) as (see Appendix)
\begin{align}
&H_{magnon}=\sum_{\bm{k},\alpha=\pm}E_{\bm{k},\alpha}b^{\dagger}_{\bm{k},\alpha}b_{\bm{k},\alpha},\notag\\
&c_{\bm{k},i}=\sum_{\alpha=\pm}\left\{[\hat{Q}_{\bm{k}}]_{i,\alpha}b_{\bm{k},\alpha}+[\hat{Q}_{\bm{k}}]_{i,\alpha+2}b^{\dagger}_{-\bm{k},\alpha}\right\},\label{magnonham}
\end{align}
where $\hat{Q}_{\bm{k}}$ and $E_{\bm{k},\alpha}$ are a paraunitary matrix and magnon eigenenergies for momentum $\bm{k}$, respectively.
We define magnon creation and annihilation operators $(b,b^\dagger)$.
The dispersion relations in the $k_x$ direction without and with the DM interaction are shown in Fig. $\ref{fig1}$(b).
The lattice constant $a$ is taken as unity.
There is a $D$-independent finite gap at $\bm{k}=0$ due to the easy-axis anisotropy.
In the presence of the DM interaction, two magnon branches ($\alpha=\pm$) have nonreciprocal band structures that are asymmetric with respect to $k_x$.
Ref. [\onlinecite{okuma}] has shown that the spin angular momentum carried by $\alpha=\pm$ magnons takes a quantized value $\pm1$.

Next, we consider coupling between the antiferromagnet and polarized light.
We generalize a theory of magnon-photon interaction in cavity spintronics \cite{cavity1,cavity2} to the case with arbitrary polarization and nonreciprocity, which is then applied to antiferromagnetic spin pumping.
The microscopic origin of this coupling is the Zeeman effect between the spin system and the alternating magnetic field of light.
By quantizing both the spin fluctuation and the light in the Zeeman coupling, we obtain the magnon-photon interaction Hamiltonian (see Appendix):
\begin{align}
H_{mp}=-\Delta(b_{\bm{k}=0,\bm{e}}^{\dagger}a_{\bm{e}}+h.c.),\label{mainresult}
\end{align}
where $\Delta$ is a coupling constant that depends on $\omega$ and $K/J$ and $a_{\bm{e}}$ is the photon annihilation operator with the polarization vector $\bm{e}=(e_1,e_2,0)^t$.
Any polarized state can be expressed as a superposition of left- and right-handed circular polarizations $\bm{e}_{\pm}=\mp(1,\pm i,0)^t/\sqrt{2}$ \cite{sakurai}, which correspond to the $z$-component spin angular momentum $\pm1$:
\begin{align}
a^{\dagger}_{\bm{e}}=\frac{-e_1+ie_2}{\sqrt{2}}a^{\dagger}_{+}+\frac{e_1+ie_2}{\sqrt{2}}a^{\dagger}_{-}.\label{linearcombination}
\end{align}
In the same manner, we define a polarized magnon state at $\bm{k}=0$:
\begin{align}
b^{\dagger}_{\bm{k}=0,\bm{e}}=\frac{-e_1+ie_2}{\sqrt{2}}b^{\dagger}_{\bm{k}=0,+}+\frac{e_1+ie_2}{\sqrt{2}}b^{\dagger}_{\bm{k}=0, -}.\label{magnonlinearcombination}
\end{align}
Since there exists a spin degeneracy at $\bm{k}=0$ even in the presence of the DM interaction\cite{anisotropyeffect}, any polarized magnon state is an energy eigenstate.

The Hamiltonian ($\ref{mainresult}$) indicates that the polarization of a photon can be converted into that of a uniform ($\bm{k}=0$) antiferromagnetic magnon without changing the polarization.
In other words, we can generate any superposition state of $S^z=\pm1$ magnon states.
In the case of ferromagnets, there is only one mode with $S^z=-1$, and only the right-handed component of the photon polarization interacts with magnons.
In contrast, for the case of antiferromagnets, there are two degenerate modes with $S^z=\pm1$, and perfect spin conservation holds in the magnon-photon interconversion process.

In addition to the main topic of spin pumping, our theory describes cavity spintronics with polarization degrees of freedom.
In the field of cavity spintronics \cite{cmhu}, a cross-discipline of spintronics and quantum information, ferromagnetic \cite{cmhu,xchang} and antiferromagnetic \cite{cavity1,cavity2} magnons in a cavity of photons have been studied in terms of the magnon-polariton because of the magnon’s long lifetime.
By using geometry in which the magnetic order is set to be parallel to the photon propagation direction, we can realize quantum states with any magnon polarization $\bm{e}$ using the corresponding polarized photon.
In addition, the uniform magnon mode with $\bar{\bm{e}}=(-ie^{*}_2,ie^{*}_1,0)^t$, which satisfies $\bar{\bm{e}}^{*}\cdot\bm{e}=0$, does not couple with the photons.
This mode can be interpreted as a ``magnon dark mode" with tunable polarization.
The magnon-photon coupling Hamiltonian can be directly applied to cavity spintronics, with a focus on the second quantization \cite{xchang, cavity1,cavity2}.

\subsection{Quantum theory of nonreciprocal spin pumping}
Let us briefly review conventional spin pumping in ferromagnets.
In ferromagnetic spin pumping, uniform magnons excited by ferromagnetic resonance are used as the source of spin current.
Because of their reciprocity, uniform magnons do not have a finite group velocity, resulting in the absence of a drift spin current.
However, excited magnons have a quantized spin angular momentum $S^z=-1$, and magnon spin accumulates throughout the whole region of the ferromagnet.
The magnon spin is converted into electron spin in the attached metal (Pt) via s-d coupling with the ferromagnet, and diffusive spin current is generated at the interface.
Refs. [\onlinecite{antiferrospinpump1,antiferrospinpump2}] have generalized this mechanism to antiferromagnets under polarized light, where the $z$-component magnon spin is $i(e_1e_2^{*}-e_1^{*}e_2)$.
Although this approach cannot avoid nonequilibrium bulk magnetization, it is an interesting possibility for antiferromagnetic spintronics.
Unfortunately, a significant signal has not yet been observed \cite{antiferrospinpump2}.

\begin{figure}[]
\begin{center}
\includegraphics[width=8cm,angle=0,clip]{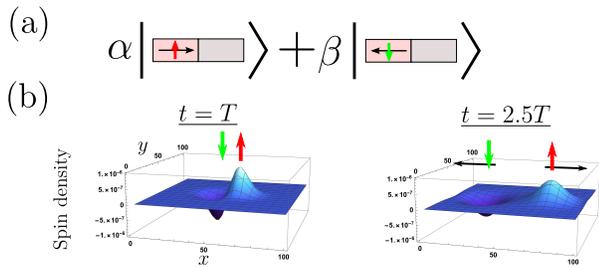}
\caption{(a) Schematic image of magnonic superposition state.
(b) Wavepacket dynamics of the magnonic superposition state for $100\times100$ sites.
The sum of the $z$-component spin $s^z_i$ for $2\times2$ sites (unit: $[/a^2]$) is plotted as a function of position for $t=T$ and $t=2.5T$. 
Pair creation of wavepackets occurs with positive and negative spin angular momentum, which propagate in opposite directions.
 }
\label{fig2}
\end{center}
\end{figure}

Here, we propose a different mechanism of spin pumping that does not induce a stray magnetic field in the bulk of an antiferromagnet.
Let us consider uniform magnons for $D\neq0$.
Owing to their nonreciprocal nature, $\alpha=\pm$ modes have finite group velocities $\pm v=\partial E_{\bm{k},\pm}/\partial k_x|_{\bm{k}=0}$.
For such states, the expectation value of the drift spin current operator 
\begin{align}
j_x^{S_z}=S^{z}_{tot}v(b^{\dagger}_{ \bm{k}=0,+}b_{ \bm{k}=0,+}-b^{\dagger}_{ \bm{k}=0,-}b_{ \bm{k}=0,-})
\end{align}
remains and does not depend on $\alpha=\pm$:
\begin{align}
\langle \bm{k}=0,\pm |j_x^{S_z}|\bm{k}=0,\pm\rangle=(\pm1)\times(\pm v)=v,
\end{align}
where $|\bm{k}=0,\alpha\rangle\equiv b^\dagger_{\bm{k},\alpha}|0\rangle$ with $|0\rangle$ being the Fock vacuum.
Thus, uniform magnons with polarization $\bm{e}$ excited through the interaction ($\ref{mainresult}$), which can be written as the superposition of $\alpha=\pm$ modes, carry a constant spin current:
\begin{align}
\langle \bm{k}=0,\bm{e} |j_x^{S_z}|\bm{k}=0,\bm{e}\rangle=v.\label{anyspincurrent}
\end{align}
Eq. ($\ref{anyspincurrent}$) shows that we can expect a constant drift spin pumping signal for any polarization.
Although there is a contribution from the diffusive spin current in the total spin pumping signal, we can extract the contribution from the drift spin current by using linearly polarized light ($S^z=0$), which does not induce bulk spin accumulation.
It is interesting to note that magnon states with linear polarization have a finite drift spin current even though they do not have a net finite group velocity due to the equal-weight superposition of states with group velocity $\pm v$ [Fig. $\ref{fig2}$(a)].
Thus, nonreciprocal spin pumping under linearly polarized light can be regarded as a pure spin current generation, like the spin Hall effect in electron systems \cite{mns,kanemele}, although the mechanism is completely different because there is no net particle flow or thermal current that is perpendicular to the spin current.

The mechanism using linearly polarized light does not require a net magnetic moment of the ground state or finite spin accumulation in the whole region of the magnet.
Thus, we can generate the spin current without generating a net magnetic field, which solves the long-standing problem of magnetic-field-free spintronics.
In addition, this new mechanism makes use of the spin current with a driving force. 
This finding indicates that nonreciprocal spin pumping is more efficient than conventional spin pumping, which has not been observed in antiferromagnets with significant signals \cite{antiferrospinpump2,pre}.
In the following, we quantitatively compare the two mechanisms via numerical simulations based on Landau-Lifshitz-Gilbert (LLG) equation.


\section{Classical theory of nonreciprocal spin pumping\label{sec3}}
Thus far, we have developed a quantum theory of nonreciprocal spin pumping in order to discuss the microscopic mechanism of polarization interconversion between one magnon and one photon. To quantitatively estimate the injected spin current in a realistic situation, however, the quantum treatment is computationally expensive.
In this section, we perform numerical simulations based on the LLG equation with an oscillating field.
Since the LLG equation is classical, the spin wave and its spin angular momentum are no longer quantized. Thus, the following simulations treat spin transport in the presence of many magnons, which is valid for a spin pumping measurement under magnetic resonance. 
To understand nonreciprocal spin pumping, it is useful to compare these complementary descriptions.

\subsection{Model and material parameters}
We treat the classical spin system described by the spin Hamiltonian ($\ref{1dantiferro}$).
We omit the $z$ direction for simplicity, while we retain the $y$ direction to keep the antiferromagnetic sublattice structure along the interface.
We perform numerical simulations based on the LLG equation for spin angular momentum under light:
\begin{align}
\frac{d\bm{s}_{\bm{i}}}{dt}=-\gamma\bm{s}_{\bm{i}}\times(\bm{H}_{eff}+\bm{h}(t))-\alpha\bm{s}_{\bm{i}}\times\frac{d\bm{s}_{\bm{i}}}{dt},\label{LLG}
\end{align}
where $\gamma$ is the gyromagnetic ratio, $\gamma\bm{H}_{eff}\equiv\partial H_{spin}/\partial\bm{S}_{\bm{i}}$, $\bm{s}_{\bm{i}}\equiv\bm{S}_{\bm{i}}/S_0$, $\bm{h}(t)$ is the oscillating magnetic field of light, and $\alpha$ is the Gilbert damping constant.

The strength of the DM interaction is strongly dependent on the setup.
To realize a nonreciprocal band structure, Ref. [\onlinecite{rcheng}] proposed a method in which a tunable DM interaction $(D/J\sim0.01)$ is induced by an electric field. 
Another example is the intrinsic DM interaction in a noncentrosymmetric antiferromagnet.
A recent neutron scattering measurement has shown the existence of a nonreciprocal magnon band structure in $\alpha$-Cu$_2$V$_2$O$_7$ with a large DM interaction ($D/J\sim1$) \cite{noncentrosym}.
In both cases, the exchange coupling $J$ and excitation energies of uniform magnon modes, which depend on $K/J$, are on the order of a few meV.
Here, we take $D/J=0.2$ and $K/J=-0.05$ as moderate parameters.
The corresponding resonant frequency is a few terahertz (THz), and we use the parameters of the THz pulse laser summarized in Appendix.

\begin{figure}[]
\begin{center}
\includegraphics[width=8cm,angle=0,clip]{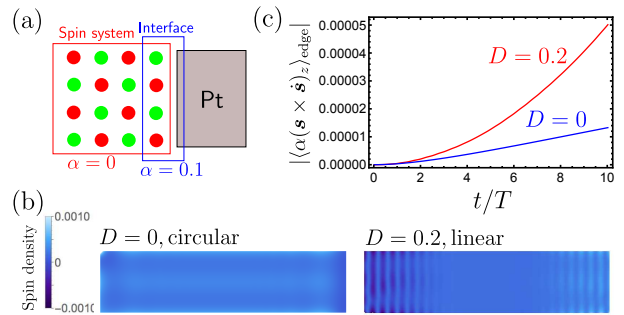}
\caption{Numerical results for $400\times100$ sites with $D=0$ and $0.2$, $J=1, K=-0.05$. (a) Schematic image of the model of the spin pumping measurement. The effect of Pt is taken as large Gilbert damping. We ignore Gilbert damping in the bulk of the spin system, which is valid for short time scales.
(b) Spin density distribution at $t=10T$.
(c) Time dependence of the spin injection rate. The value is calculated as the site average of the Gilbert damping term at the right edge (unit: $[/a^2T]$).
 }
\label{fig3}
\end{center}
\end{figure}

\subsection{Numerical results}
To get the feel of the magnonic superposition state, we first explore the dynamics of a small wavepacket before discussing the spin pumping measurement.
Let us consider a position-dependent oscillating field with one period:
\begin{align}
\bm{H}(\bm{x},t)=\theta(T-t)\bm{h}(t)\exp\left(-\frac{x^2+y^2}{2(10a)^2} \right),
\end{align}
where $\theta$ is a step function, $T$ is the period of light, and $a$ ($\sim$1 nm) is a lattice constant \cite{diffraction}.
Because we consider a short time scale ($\sim\mathcal{O}(T)$), we set $\alpha=0$, which is valid for a magnetic insulator with low dissipation.
The $z$-component spin density dynamics for $100\times100$ sites under linearly polarized light is shown in Fig. $\ref{fig2}$(b).
The wavepacket with positive spin density propagates in the positive $x$ direction, while the other wavepacket propagates in the negative $x$ direction, which induces a $z$-component spin current in the $x$ direction.
This result classically reproduces the discussions in quantum theory.

Next, we simulate nonreciprocal spin pumping.
In an actual spin pumping measurement, the alternating magnetic field is uniform over the range of 100 $\mu$m.
Here, we solve the LLG equation for $400\times100$ sites under a position-independent oscillating field with ten periods, which corresponds to a short-pulse laser.
The effect of Pt is taken into account by setting the large Gilbert damping $\alpha=0.1$ at the right edge of the system [Fig. $\ref{fig3}$(a)]. 
We again ignore damping in the bulk of the antiferromagnet.
For a rigorous treatment, the electron spin and charge transport in Pt should also be considered, although it is computationally more expensive. 
We leave this problem as future work.

For comparison, both reciprocal ($D=0$, left-handed circular polarization) and nonreciprocal ($D=0.2$, linear polarization) spin pumping measurements are considered.
The spin density at $t=10T$ is shown in Fig. $\ref{fig3}$(b).
In the case of reciprocal spin pumping, the spin angular momentum of the circularly polarized light is converted into that of a uniform magnon, and spin accumulates throughout the whole region of the antiferromagnet.
In the case of nonreciprocal spin pumping, the linearly polarized light has no net spin angular momentum, and there is no net spin accumulation in the antiferromagnet.
Instead, the positive spin accumulates near the right edge, while the negative spin accumulates near the left edge. 
This phenomenon is the consequence of the nonreciprocal spin current in the magnonic superposition state.
Note that spin oscillation is observed in the case of nonreciprocal spin pumping.
This spin oscillation is a precursory phenomenon of the magnon Bose-Einstein condensate, which will be discussed later.
The slight spin oscillation in the $y$ direction, which is observed in both cases, is a finite size effect. 

To evaluate the spin current injected into Pt, we calculate the time dependence of the Gilbert damping term at the right edge [Fig. $\ref{fig3}$(c)], which describes the spin angular momentum transfer over time from the antiferromagnet to the metal.
As shown in Fig. $\ref{fig3}$(c), the nonreciprocal spin current for realistic parameters can be much larger than the conventional spin current after a sufficient amount of time. This result can be qualitatively interpreted as follows.
In the case of reciprocal spin pumping, the spin angular momentum of one photon is converted into spin accumulation for the whole region.
In the case of nonreciprocal spin pumping, the large amount of spin angular momentum created by one photon accumulates near the left and right edges. Thus, the nonequilibrium spin density at the interface is much larger than that in the case of reciprocal spin pumping.

The remaining issue is spin oscillation near the boundary, which cannot be understood in a linear approximation.
In the following, we relate this aspect to the two-component Bose-Einstein condensate.

\begin{figure}[]
\begin{center}
\includegraphics[width=7cm,angle=0,clip]{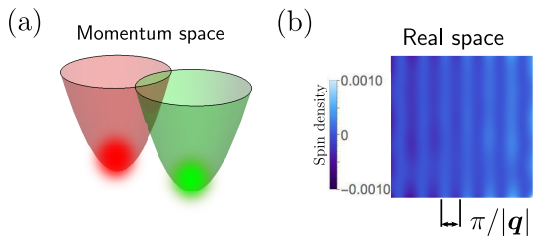}
\caption{Stripe superfluid phase in a two-component magnon Bose-Einstein condensate.
(a) Schematic image of the magnon distribution in momentum space.
(b) Spin density distribution at $t=10T$ for $100\times100$ sites with $\alpha=0$ at the right edge.
 }
\label{fig4}
\end{center}
\end{figure}

\section{Breakdown of the superposition state\label{sec4}}
In the following, we discuss the analogous features of nonreciprocal magnons in the antiferromagnet and ultracold atomic systems with spin-orbit interaction.
The spin oscillation observed in the numerical simulation can be understood by generalizing the magnon Bose-Einstein condensate \cite{magnonbec} to the antiferromagnetic nonreciprocal band structure.
At the edges of the antiferromagnet, the physics of nonreciprocal spin pumping cannot be well described by only uniform magnons because of the strong magnon-magnon scatterings, including higher-order interactions.
High-energy magnons are scattered into low-energy states via these interactions and finally remain at the bottoms of bands.
In the nonreciprocal magnon band structure, potential minima are located at two finite momenta:
\begin{align}
\bm{k}=\pm\left(\arctan\left[\frac{D}{J}\right],0,0\right)\equiv \pm\bm{q},\label{ossmomentum}
\end{align}
which are solutions of $\partial E/\partial k_x=0$.
Ideally, a two-spin-component magnon Bose-Einstein condensate occurs at $\bm{k}=\pm \bm{q}$, with the ratio determined by the polarization [Fig. \ref{fig4}(a)].
This is a magnonic analogue of the spin-orbit-coupled Bose-Einstein condensate, which has been extensively studied in ultracold atomic systems \cite{goldman}.
When $S^z=\pm1$ bosons condense with equal weight, the spin density oscillates in real space \cite{goldman}:
\begin{align}
\rho_{S^z}(\bm{r})\propto\cos(2\bm{q}\cdot \bm{r}).\label{spinoscillation}
\end{align}
This phase is known as a ``stripe superfluid" or ``standing wave phase" [Fig. $\ref{fig4}$].
The absence of spin density in the ground state of antiferromagnets would facilitate the detection of this spin oscillation.

The spin accumulation with oscillation under linearly polarized light can be understood by this stripe superfluid phase.
After a sufficient amount of time for a small sample, the spin accumulation near the right (left) edge reaches the opposite side of the sample, and we can expect a spin stripe structure described by Eq. ($\ref{spinoscillation}$).
Let us consider the case of $100\times100$ sites without Pt ($\alpha=0$).
Fig. $\ref{fig4}$(b) shows the existence of the stripe superfluid phase.
The period of the observed spin oscillation is consistent with the value of $15.9\ a$ calculated by Eq. ($\ref{spinoscillation}$).


\section{Summary}
We have proposed a theory of the magnonic superposition state with nonreciprocal current.
In the first half of this paper, we developed a quantum theory of polarization interconversion between one magnon and one photon.
In particular, by making use of nonreciprocal magnon bands and linearly polarized photons, we proposed nonreciprocal spin current generation without generating a net magnetic field, which solves the long-standing problem of magnetic-field-free spintronics.
In the second half of this paper, we performed numerical simulations based on the LLG equation, in order to quantitatively evaluate nonreciprocal spin pumping.
We find that for realistic parameters, the spin current can be injected more efficiently than the conventional current.
We also find that the spin oscillation observed near the edges is a signature of the stripe superfluid phase in the two-component Bose-Einstein condensate.

\section*{Acknowledgements} I acknowledge many fruitful discussions with Hideaki Maebashi, Shuichi Murakami, Akihiro Okamoto, Masahiro Sato, and Tomonari Mizoguchi.
I am grateful to Masaya Kunimi for introducing a relevant paper.
I was supported by the Japan Society for the Promotion of Science KAKENHI (Grant No. 16J07110, 18J01610).

\appendix
\section{Quantization of Zeeman coupling between light and an antiferromagnet}
Here, we microscopically derive the magnon-photon coupling Hamiltonian.
The microscopic origin of this coupling is the Zeeman effect between the spin system and the magnetic field $\bm{\mathcal{B}}$ created by photons:
\begin{align}
H_{Z}=g\mu_B\bm{\mathcal{B}}\cdot\sum_{\bm{i}}\bm{S}_{\bm{i}}
=\frac{g\mu_B}{2}\left(\mathcal{B}^+S^-_{tot}+\mathcal{B}^-S^+_{tot}\right).\label{zeeman}
\end{align}
Let us quantize the magnetic field of light and the spin ladder operators.
The magnetic field is given in terms of photon annihilation and creation operators ($a,a^{\dagger}$):
\begin{align}
\bm{\mathcal{B}}=i\beta\sqrt{\omega}\hat{z}\times(\bm{e}a_{\bm{e}}-\bm{e}^*a^{\dagger}_{\bm{e}}),\label{photon}
\end{align}
where $\beta$ is a constant and $\omega$ is the frequency of the photon.
The total spin ladder operator $S^{+}_{tot}$ is given in terms of magnon modes by
\begin{align}
S^{+}_{tot}&=\sum_{\bm{R}}\left(S^{+}_{\bm{R},A}+S^{+}_{\bm{R},B}\right)=\sqrt{2S_0L^3}\left(c_{\bm{k}=0,A}+c^{\dagger}_{\bm{k}=0,B}\right)\notag\\
&=F(-K/3J)\left(b_{\bm{k}=0,-}+b^{\dagger}_{\bm{k}=0,+}\right),\label{spinladder}
\end{align}
where $L$ is the size of the magnet and $F(x)=2\sqrt{2S_0L^3}(1+x)^{3/2}\{(1+x)^2-1\}^{-1/4}$.
We have used the explicit form of $\hat{Q}_{\bm{k}=0}$.
Substituting Eqs. (\ref{photon}, and \ref{spinladder}) and Eqs. (\ref{linearcombination}, and \ref{magnonlinearcombination}), we obtain
\begin{align}
H_{mp}=-\frac{g\mu_B}{\sqrt{2}}\beta\sqrt{\omega}F(-K/3J)(b_{\bm{k}=0,\bm{e}}^{\dagger}a_{\bm{e}}+h.c.),
\end{align}
where we have used the rotating wave approximation ($ab=a^{\dagger}b^{\dagger}=0$).

\section{Derivation of the classical ground state}
Here, we show that the classical ground state of the spin Hamiltonian given in the main text is the N\'{e}el state for large $|K|$.
For this purpose, we use the Luttinger-Tisza method \cite{luttinger}.
The classical total energy of the spin system is given in terms of the Fourier transform of spins as
\begin{align}
E_{tot}=\sum_{\bm{q}}\bm{S}^T_{-\bm{q}}H_{\bm{q}}\bm{S}_{\bm{q}},
\end{align}
where 
\begin{widetext}
\begin{align}
H_{\bm{q}}=
\begin{pmatrix}
J(\cos q_x+\cos q_y+\cos q_z)&i D\sin q_x&0\\
-iD\sin q_x&J(\cos q_x+\cos q_y+\cos q_z)&0\\
0&0&J(\cos q_x+\cos q_y+\cos q_z)+K
\end{pmatrix}.
\end{align}
\end{widetext}
According to the Luttinger-Tisza framework, the eigenvector of $H_{\bm{q}}$ with the smallest eigenvalue corresponds to the classical ground state if it satisfies the local constraints:
\begin{align}
|\bm{S}_{\bm{i}}|^2=1,
\end{align}
where $S_0=1$.
The eigenvectors and their eigenvalues take the following two forms:
\begin{align}
&\begin{pmatrix}
\alpha\\
\beta\\
0
\end{pmatrix}
\mathrm{with\ }\epsilon_1(\bm{q})=J(\cos q_x+\cos q_y+\cos q_z)\pm |D\sin q_x|,\notag\\
&\begin{pmatrix}
0\\
0\\
1
\end{pmatrix}
\mathrm{with\ }\epsilon_2(\bm{q})=J(\cos q_x+\cos q_y+\cos q_z)+K.
\end{align}
Thus, for sufficiently small $K<0$, the latter eigenvector with $q_x=q_y=q_z=\pi$ has the smallest eigenvalue. This state is simply the N\'{e}el state.

\section{Bogoliubov transformation of the magnon Hamiltonian}
Here, we present a theory of the eigenenergy problem of magnon modes.
We start with the magnon Hamiltonian, which is obtained by substituting Eq. (\ref{holstein}) into Eq. (\ref{1dantiferro}):
\begin{align}
H_{magnon}&=\frac{1}{2}\sum_{\bm{k}}\Psi^{\dagger}_{\bm{k}}\hat{H}_{\bm{k}}\Psi_{\bm{k}},\notag\\
\hat{H}_{\bm{k}}&=
\begin{pmatrix}
X(\bm{k}) & 0 & 0 & Y_{-}(\bm{k})\\
0 & X(\bm{k}) & Y_{+}(\bm{k})& 0\\
0& Y_{+}(\bm{k})&X(\bm{k}) & 0\\
Y_{-}(\bm{k})&0 &0 &X(\bm{k})
\end{pmatrix},\label{hbdg}
\end{align}
where $\Psi^{\dagger}_{\bm{k}}=(c^{\dagger}_{\bm{k},A},c^{\dagger}_{\bm{k},B},c_{-\bm{k},A},c_{-\bm{k},B})$, $\hat{H}_{\bm{k}}$ is a bosonic Bogoliubov-de Gennes (BdG) Hamiltonian, $X(\bm{k})=2S_0(3J-K)$, and $Y_{\pm}(\bm{k})=-2S_0(J\cos k_x+J\cos k_y+J\cos k_z\mp D\sin k_x)$. 
In general, the magnon eigenenergy problem cannot be solved by simply diagonalizing the quadratic matrix with unitary matrices because a naive unitary transformation in the presence of $cc,c^\dagger c^\dagger$ breaks the bosonic commutation relation.
Instead, the eigenstates and eigenenergies of $H$ can be obtained by the bosonic Bogoliubov transformation
\cite{colpa,maestro}:
\begin{align}
&\hat{Q}^{\dagger}_{\bm{k}}\hat{\mathcal{H}}_{\bm{k}}\hat{Q}_{\bm{k}}=
\begin{pmatrix}
\hat{E}_{\bm{k}} & 0\\
0 & \hat{E}_{-\bm{k}}
\end{pmatrix},
\end{align}
where $\hat{Q}$ is the paraunitary matrix in Eq. (\ref{magnonham}) and $\hat{E}_{\bm{k}}$ is a diagonal matrix whose diagonal elements are magnon eigenenergies.
Explicit forms of $\hat{Q}$ and $\hat{E}_{\bm{k}}$ are given in the following section.

\section{Derivation of the energy spectrum and paraunitary matrix}
We here solve the bosonic Bogoliubov-de Gennes (BdG) Hamiltonian analytically by using the bosonic Bogoliubov transformation.

The general form of the bosonic BdG Hamiltonian is given by
\begin{align}
\hat{\mathcal{H}}_{\bm{k}}=
\begin{pmatrix}
\hat{A}_{\bm{k}} & \hat{B}_{\bm{k}} \\
\hat{B}^{*}_{-\bm{k}}  & \hat{A}^{*}_{-\bm{k}}
\end{pmatrix},\label{bdgmatrix}
\end{align}
where $\hat{A}$ is a $N\times N$ Hermitian matrix and $\hat{B}$ a $N\times N$ matrix.
In the following, we assume that $\hat{\mathcal{H}}_{\bm{k}}$ is positive definite.
The bosonic Bogoliubov transformation is defined as
\begin{align}
&\hat{Q}^{\dagger}_{\bm{k}}\hat{\mathcal{H}}_{\bm{k}}\hat{Q}_{\bm{k}}=
\begin{pmatrix}
\hat{E}_{\bm{k}} & 0\\
0 & \hat{E}_{-\bm{k}}
\end{pmatrix},\label{positivediag}
\end{align}
where the paraunitary matrix $\hat{Q}_{\bm{k}}$ satisfies
\begin{align}
\hat{Q}_{\bm{k}}^{\dagger}\hat{\Sigma}_3\hat{Q}_{\bm{k}}=\hat{Q}_{\bm{k}}\hat{\Sigma}_3\hat{Q}_{\bm{k}}^{\dagger}=\hat{\Sigma}_3,\label{paraunitarycondition}
\end{align}
where $[\hat{\Sigma}_3]_{i,j}=\delta_{ij}\sigma_j$ with $\sigma_j=+1$ for $j=1,\cdots, N$ and $\sigma_j=-1$ for $j=N+1,\cdots, 2N$.
All we have to do is determine $\hat{Q},\hat{Q}^{\dagger}$ satisfying Eqs. (\ref{positivediag}) and (\ref{paraunitarycondition}).

For positive definite Hermitian matrix $\hat{\mathcal{H}}_{\bm{k}}$, we can perform the Cholesky decomposition
\begin{align}
\hat{\mathcal{H}}_{\bm{k}}=\hat{K}^{\dagger}_{\bm{k}}\hat{K}_{\bm{k}},
\end{align}
where $\hat{K}_{\bm{k}}$ is an upper triangle matrix.
Using $\hat{K}_{\bm{k}}$ and $\hat{K}^{\dagger}_{\bm{k}}$, we define a unitary matrix
\begin{align}
\hat{U}_{\bm{k}}\equiv\hat{K}_{\bm{k}}\hat{Q}_{\bm{k}}
\begin{pmatrix}
\hat{E}_{\bm{k}}^{-\frac{1}{2}} & 0\\
0 & \hat{E}^{-\frac{1}{2}}_{-\bm{k}}
\end{pmatrix}\label{uni}
\end{align}
and the dual Hamiltonian
\begin{align}
\hat{\mathcal{H}'}_{\bm{k}}\equiv \hat{K}_{\bm{k}}\hat{\Sigma}_3\hat{K}^{\dagger}_{\bm{k}},\label{dualham}
\end{align}
which is a Hermitian matrix.
Naturally, the dual Hamiltonian (\ref{dualham}) is diagonalized by the unitary matrix (\ref{uni}):
\begin{align}
&\hat{U}_{\bm{k}}^{\dagger}\hat{\mathcal{H}'}_{\bm{k}}\hat{U}_{\bm{k}}\notag\\
&=
\begin{pmatrix}
\hat{E}_{\bm{k}}^{-\frac{1}{2}} & 0\\
0 & \hat{E}^{-\frac{1}{2}}_{-\bm{k}}
\end{pmatrix}
\hat{Q}_{\bm{k}}^{\dagger}\hat{K}_{\bm{k}}^{\dagger} \hat{K}_{\bm{k}}\hat{\Sigma}_3\hat{K}^{\dagger}_{\bm{k}}  \hat{K}_{\bm{k}}\hat{Q}_{\bm{k}}
\begin{pmatrix}
\hat{E}_{\bm{k}}^{-\frac{1}{2}} & 0\\
0 & \hat{E}^{-\frac{1}{2}}_{-\bm{k}}
\end{pmatrix}\notag\\
&=
\begin{pmatrix}
\hat{E}_{\bm{k}} & 0\\
0 & -\hat{E}_{-\bm{k}}
\end{pmatrix}.
\end{align}
Thus, we can obtain the magnon eigenvalues by diagonalizing the Hermitian matrix (\ref{dualham}). 
After determining $\hat{E}_{\bm{k}}$ and $\hat{E}_{-\bm{k}}$ by the diagonalization, we can determine the paraunitary matrices as
\begin{align}
\hat{Q}_{\bm{k}}\equiv\hat{K}_{\bm{k}}^{-1}\hat{U}_{\bm{k}}
\begin{pmatrix}
\hat{E}_{\bm{k}}^{\frac{1}{2}} & 0\\
0 & \hat{E}^{\frac{1}{2}}_{-\bm{k}}
\end{pmatrix}.
\end{align}

Using the above method, we here give the explicit forms of $\hat{K}$, $\hat{K}^{-1}$, $\hat{\mathcal{H}}'$, and $\hat{U}$ for the reciprocal and nonreciprocal magnon Hamiltonians (\ref{magnonham}).
The upper triangle matrix $\hat{K}_{\bm{k}}$ in the Cholesky decomposition is given by
\begin{align}
\hat{K}=\sqrt{\frac{1}{X}}
\begin{pmatrix}
X & 0 & 0 & Y_{-}\\
0 & X & Y_{+}& 0\\
0&0&\sqrt{X^2-Y_{+}^2 }& 0\\
0&0 &0 &\sqrt{X^2-Y_{-}^2 }
\end{pmatrix},\label{triangleexpression}
\end{align}
and its inverse is
\begin{align}
\hat{K}^{-1}=\sqrt{\frac{1}{X}}
\begin{pmatrix}
1 & 0 & 0 & -Y_{-}/\sqrt{ X^2-Y_{-}^2  }\\
0 & 1 & -Y_{+}/\sqrt{ X^2-Y_{+}^2  }& 0\\
0&0&X/\sqrt{ X^2-Y_{+}^2  }& 0\\
0&0 &0 &X/\sqrt{X^2-Y_{-}^2 }
\end{pmatrix},
\end{align}
where we omit $(\bm{k})$ for simplicity.
Using Eq. (\ref{triangleexpression}), we obtain the dual Hamiltonian
\begin{widetext}
\begin{align}
\hat{\mathcal{H}}'=\frac{1}{X}
\begin{pmatrix}
X^2-Y_{-}^2 & 0 & 0 & -Y_{-}\sqrt{ X^2-Y_{-}^2  }\\
0 & X^2-Y_{+}^2   & -Y_{+}\sqrt{ X^2-Y_{+}^2  }& 0\\
0& -Y_{+}\sqrt{ X^2-Y_{+}^2  }&-(X^2-Y_{+}^2  )& 0\\
 -Y_{-}\sqrt{ X^2-Y_{-}^2  }&0 &0 &-(X^2-Y_{-}^2  )
\end{pmatrix}.
\end{align}
\end{widetext}
This Hamiltonian can be diagonalized by using a unitary matrix
\begin{widetext}
\begin{align}
\hat{U}=
\begin{pmatrix}
0 & -\frac{(\sqrt{ X^2-Y_{-}^2  }+X) }{Y_{-}}& -\frac{(\sqrt{ X^2-Y_{-}^2  }-X)}{Y_{-}} & 0\\
-\frac{(\sqrt{ X^2-Y_{+}^2  }+X) }{Y_{+}} & 0  &0& -\frac{(\sqrt{ X^2-Y_{+}^2  }-X) }{Y_{+}}\\
1&0&0& 1\\
0&1&1&0
\end{pmatrix},
\end{align}
\end{widetext}
and the magnon eigenvalues are given by
\begin{align}
E_{\bm{k},\pm}=\sqrt{X^2(\bm{k})-Y^2_{\pm}(\bm{k})}.\label{1dmagdispersion}
\end{align}
The corresponding paraunitary matrix is given by
\begin{widetext}
\begin{align}
\hat{Q}_{\bm{k}}&\equiv\hat{K}_{\bm{k}}^{-1}\hat{U}_{\bm{k}}
\begin{pmatrix}
\hat{E}_{\bm{k}}^{\frac{1}{2}} & 0\\
0 & \hat{E}^{\frac{1}{2}}_{-\bm{k}}
\end{pmatrix}\notag\\
&=\begin{pmatrix}
0 & -\frac{\sqrt{X}(\sqrt{ X^2-Y_{-}^2  }+X) }{Y_{-}(X^2-Y_{-}^2)^{1/4} }&\frac{\sqrt{X}(\sqrt{ X^2-Y_{-}^2  }-X) }{Y_{-}(X^2-Y_{-}^2)^{1/4} }& 0\\
 -\frac{\sqrt{X}(\sqrt{ X^2-Y_{+}^2  }+X) }{Y_{+}(X^2-Y_{+}^2)^{1/4} }& 0  &0& \frac{\sqrt{X}(\sqrt{ X^2-Y_{+}^2  }-X) }{Y_{+}(X^2-Y_{+}^2)^{1/4} }\\
\frac{\sqrt{X} }{(X^2-Y_{+}^2)^{1/4} }&0&0& \frac{\sqrt{X} }{(X^2-Y_{+}^2)^{1/4} }\\
0&\frac{\sqrt{X} }{(X^2-Y_{-}^2)^{1/4} }&\frac{\sqrt{X} }{(X^2-Y_{-}^2)^{1/4} }&0
\end{pmatrix},\label{1dqmat}
\end{align}
\end{widetext}
where we have used $Y_{+}(k)=Y_{-}(-k)$.
Note that reciprocal and nonreciprocal magnon systems have the same $\hat{Q}$ at $\bm{k}=0$.

\section{Numerical conditions of the LLG equation}
Here, we present the conditions of the numerical simulations.
In the following, the unit of energy is taken as $JS_0$, where $J$ is typically a few meV.
We have considered alternating magnetic fields $\gamma\bm{h}(t)=0.001(\cos\omega_{res}t,\sin\omega_{res}t,0)/\sqrt{2}$ for the reciprocal case under circularly polarized light and $\gamma\bm{h}(t)=0.001(0,\cos\omega_{res}t,0)$ for the nonreciprocal case under linearly polarized light.
Here, $\omega_{res}=2S_0\sqrt{(2J-K)^2-(2J)^2}$ is the resonant frequency for two dimensions.
The resonant frequency and the magnetic field of light for the above parameters correspond to $\mathcal{O}(1)$ THz and $\mathcal{O}(100)$ [kA/m], respectively. Thus, the above parameters correspond to a typical THz pulse \cite{typicalpulse}.
We have solved Eq. (\ref{LLG}) with the free boundary condition by using the fourth-order Runge-Kutta method with $\Delta t=0.1/\omega_{res}$.

\section{Effects in realistic materials}
In realistic magnets, there exists magnetic domains.
Since the DM vector, which is independent of the magnetic order, determines the direction of drift spin current, contributions from such domains do not cancel out each other. 
In the case of the lattice domains that flip the direction of the DM interaction, on the other hand, the spin current flows in the opposite direction.
Thus, the sample quality affects the nonreciprocal spin pumping in the macroscopic samples.
Also, if the photon propagation direction is slightly different from the $z$ direction, the spin conservation is slightly broken due to the absence of spin rotation symmetry around the $z$ axis. In such a case, the weights of up and down magnon states in the excited superposition states are changed. In realistic systems, there also exist a lot of processes of the magnon relaxation. Unfortunately, such effects are complicated and usually cannot be discussed without the aid of phenomenological parameters. In particular, the relaxation of antiferromagnetic magnons is not well known. We leave this problem as future work.


\begin{thebibliography}{9}
\bibitem{magnonics}V. V. Kruglyak, S. O. Demokritov, and D. Grundler, J. Phys. D $\bm{43}$, 264001 (2010).
\bibitem{chumak}A. V. Chumak, V. I. Vasyuchka, A. A. Serga, and B. Hillebrands, Nature Phys. $\bm{11}$, 453 (2015).
\bibitem{cornelissen} L. J. Cornelissen, K. J. H. Peters, G. E. W. Bauer, R. A. Duine, and B. J. van Wees, Phys. Rev. B $\bm{94}$, 014412 (2016).
\bibitem{jungwirth}T. Jungwirth, X. Marti, P. Wadley, and J. Wunderlich, Nat. Nanotechnol.$\bm{11}$, 231 (2016).
\bibitem{optospintronics}P. N\v{e}mec, M. Fiebig, T. Kampfrath, and A. V. Kimel, Nat. Phys. $\bm{14}$, 229 (2018). 
\bibitem{rcheng}R. Cheng, M. W. Daniels, J.-G. Zhu, and D. Xiao, Sci. Rep. $\bm{6}$, 24223 (2016).
\bibitem{antiferrospinpump1}R. Cheng, J. Xiao, Q. Niu, and A. Brataas, Phys. Rev. Lett. $\bm{113}$, 057601 (2014).
\bibitem{igor}I. Proskurin, R. L. Stamps, A. S. Ovchinnikov, and J. I. Kishine, Phys. Rev. Lett. $\bm{119}$, 177202 (2017).
\bibitem{tsatoh}T. Satoh, S.-J. Cho, R. Iida, T. Shimura, K. Kuroda, H. Ueda, Y. Ueda, B. A. Ivanov, F. Nori, and M. Fiebig, Phys. Rev. Lett. $\bm{105}$, 077402 (2010).

\bibitem{ferri}
Y. Ohnuma, H. Adachi, E. Saitoh, and S. Maekawa, Phys. Rev. B $\bm{87}$, 014423 (2013).


\bibitem{fujimoto}S. Fujimoto, Phys.Rev.Lett. $\bm{103}$, 047203 (2009).
\bibitem{katsura}H. Katsura, N. Nagaosa, and P. A. Lee, Phys.Rev.Lett. $\bm{104}$, 066403 (2010).
\bibitem{onose}Y. Onose, T. Ideue, H. Katsura, Y. Shiomi, N. Nagaosa, and Y. Tokura, Science $\bm{329}$, 297 (2010).
\bibitem{owerre3}S. A. Owerre, J. Phys.:  Condens. Matter $\bm{29}$, 03LT01 (2017).

\bibitem{okuma}N. Okuma, Phys. Rev. Lett. $\bm{119}$, 107205 (2017).
\bibitem{kawano} M. Kawano, Y. Onose, C. Hotta, arXiv:1805.03925.

\bibitem{shindou} R. Shindou, R. Matsumoto, S. Murakami, and J. I. Ohe, Phys. Rev. B $\bm{87}$, 174427 (2013).
\bibitem{lzhang}L. Zhang, J. Ren, J. -S. Wang, and B. Li, Phys. Rev. B $\bm{87}$, 144101 (2013).
\bibitem{chisnell}R. Chisnell, J. S. Helton, D. E. Freedman, D. K. Singh, R. I. Bewley, D. G. Nocera, and Y. S. Lee, Phys. Rev. Lett. $\bm{115}$, 147201 (2015).
\bibitem{owerre} S. A. Owerre, Journal  of  Applied  Physics $\bm{120}$,  043903 (2016).

\bibitem{fyli}F. Y. Li, Y. D. Li, Y. B. Kim, L. Balents, Y. Yu, and G. Chen, Nature Commun. $\bm{7}$, 12691 (2016).
\bibitem{jianandnie}S.-K. Jian and W. Nie, Phys. Rev. B $\bm{97}$, 115162 (2018).
\bibitem{fransson}J. Fransson, A. M. Black-Schaffer, and A. V. Balatsky, Phys. Rev. B $\bm{94}$, 075401 (2016).
\bibitem{pershoguba}S. S. Pershoguba, S. Banerjee, J. C. Lashley, J. Park, H. Agren, G. Aeppli, and A. V. Balatsky, Phys. Rev. X $\bm{8}$, 011010 (2018).


\bibitem{nodalloop}S. A. Owerre, Sci. Rep.$\bm{7}$, 6931 (2017).
\bibitem{nodalloop3}S. A. Owerre, Europhys. Lett. $\bm{120}$, 57002 (2017).
\bibitem{nodalloop2}S. A. Owerre, arXiv:1801.03498.
\bibitem{ferrononreciprocal}
Y. Iguchi, S. Uemura, K. Ueno, and Y. Onose, Phys. Rev. B $\bm{92}$, 184419 (2015).
\bibitem{hayami}S. Hayami, H. Kusunose,  and Y. Motome, J. Phys. Soc. Jpn $\bm{85}$,053705 (2016).

\bibitem{noncentrosym} G. Gitgeatpong, Y. Zhao, P. Piyawongwatthana, Y. Qiu, L. W. Harriger, N. P. Butch, T. J. Sato, and K. Matan, Phys. Rev. Lett. $\bm{119}$, 047201 (2017).
\bibitem{takashima}Y.  Shiomi,   R.  Takashima,   D.  Okuyama,   G.  Gitgeatpong,P. Piyawongwatthana, K. Matan, T. J. Sato,  and E. Saitoh, Phys. Rev. B $\bm{96}$, 180414 (2017).


\bibitem{orientation}
The summation in the first term is taken over the sets of nearest-neighbours $\langle \bm{i},\bm{j}\rangle$, where $i_a\leq j_a\ (a=x,y,z)$.







\bibitem{cavity1}
H. Y. Yuan  and  X.  R.  Wang,  Appl.  Phys.  Lett. $\bm{110}$, 082403 (2017).
\bibitem{cavity2}
\O. Johansen and A. Brataas, Phys. Rev. Lett. $\bm{121}$, 087204 (2018).




\bibitem{sakurai}J. J. Sakurai, $Advanced$ $Quantum$ $Mechanics$ (Pearson Education, India, 1967).


\bibitem{anisotropyeffect} 
This degeneracy is robust against other perturbations as shown in Ref. [\onlinecite{noncentrosym}].


\bibitem{cmhu}C.-M. Hu, Physics in Canada, $\bm{72}$, No. 2, 76 (2016).
\bibitem{xchang} X. Zhang, C. -L. Zou, N. Zhu, F. Marquardt, L. Jiang and H. X. Tang, Nat. Commun. $\bm{6}$, 8914 (2015).




\bibitem{antiferrospinpump2} \O. Johansen and A. Brataas, Phys. Rev. B $\bm{95}$, 220408(R) (2017).







\bibitem{mns}Murakami, N. Nagaosa, and S. C. Zhang, Science $\bm{301}$,1348 (2003).
\bibitem{kanemele}C. L. Kane and E. J. Mele, Phys. Rev. Lett. $\bm{95}$, 226801 (2005).


\bibitem{pre}M. P. Ross, Spin dynamics in an antiferromagnet, Ph.D. thesis, Technische Universit\"{a}t M\"{u}nchen, 2013.








\bibitem{diffraction}
The simulation for such a small spot ($\sim10$nm) corresponds to a thought experiment rather than an actual experiment because of the diffraction limit ($\sim100\mu$m) of the teraherz laser.
However, this is at least not unphysical because recent developments in plasmonics enable us to create a spot that is much smaller than the diffraction limit. 




\bibitem{magnonbec}
S. O. Demokritov, V. E. Demidov, O. Dzyapko, G. A. Melkov, A. A. Serga, B. Hillebrands and A. N. Slavin, Nature $\bm{443}$, 430 (2006).

\bibitem{goldman}
N. Goldman, G. Juzeliunas, P. Ohberg, and I. B. Spielman, Rep. Prog. Phys. $\bm{77}$, 126401 (2014).




\bibitem{luttinger}
 J. M. Luttinger and L. Tisza, Phys. Rev. $\bm{70}$, 954 (1946).



\bibitem{colpa} J. H. P. Colpa, Physica A $\bm{93}$, 327 (1978).
\bibitem{maestro}
A. D. Maestro and M. Gingras, J. Phys. Cond. Matt. $\bm{16}$, 3399 (2004).
\bibitem{typicalpulse}
Y. Mukai, H. Hirori, T. Yamamoto, H. Kageyama, and K. Tanaka, Appl. Phys. Lett. $\bm{105}$, 022410 (2014).



\end{thebibliography}
\end{document}